\documentclass[sigplan]{acmart}
\renewcommand\footnotetextcopyrightpermission[1]{}
\usepackage{enumitem}
\usepackage{tikz}
\usetikzlibrary{arrows.meta,positioning}

\usepackage{xargs}  

\newcommandx{\myrightarrow}[1]{
  \xrightarrow{ \raisebox{-1.5pt}[2pt][0pt]{ \ensuremath{ \scriptstyle{#1} }}}
}

\newcommand{\canCommit}{\ensuremath{\mathit{canCommit}}}
\newcommand{\vShift}{\ensuremath{\mathit{vShift}}}

\setcopyright{none}
\acmConference[SCCP 2026]{1st Symposium on Checking Consistency Principles}{September 2026}{Boston, MA, USA}
\acmDOI{}
\acmISBN{}
\acmYear{2026}
\begin{document}

\title{Reduce Once, Verify Many: Verifying Isolation Guarantees via Hierarchical Abstractions}
\titlenote{Accepted for presentation at the 1st Symposium on Checking Consistency Principles (SCCP 2026), co-located with VLDB 2026.}

% TODO: verify author order, affiliations, and emails before submission.
\author{Shabnam Ghasemirad}
\affiliation{\institution{ETH Zurich, Switzerland}\city{}\country{}}
\email{shabnam.ghasemirad@inf.ethz.ch}

\author{Christoph Sprenger}
\affiliation{\institution{ETH Zurich, Switzerland}\city{}\country{}}
\email{sprenger@inf.ethz.ch}

\author{Si Liu}
\affiliation{\institution{Texas A\&M University, USA}\city{}\country{}}
\email{si.liu@tamu.edu}

\author{David Basin}
\affiliation{\institution{ETH Zurich, Switzerland}\city{}\country{}}
\email{basin@inf.ethz.ch}

\renewcommand{\shortauthors}{Ghasemirad et al.}

\begin{abstract}
\looseness=-1
We present a mathematically rigorous, systematic approach for the verification of database isolation guarantees, which (i) supports a spectrum of seven isolation levels, (ii) uncovers a fundamental dichotomy among isolation levels: stronger levels can be verified via refinement alone, whereas weaker levels additionally require reduction, and (iii) provides a hierarchy of abstract models that substantially simplifies proofs by factoring out their most labor-intensive parts. In particular, we eliminate the need for per-protocol reduction proofs for the weaker class of isolation levels by performing a once-and-for-all reduction at a high level of abstraction in our hierarchy. To achieve this, we develop and apply a generic theory of reduction, which is also of more general interest. Overall, our approach minimizes the user's proof effort to a single, simpler refinement of the most concrete model in our hierarchy. All our results are formalized in Isabelle/HOL.
\end{abstract}

% ACM CCS Concepts %TODO: change

%TODO: check
%nobi: no need, to save space
%\keywords{concurrency control protocols, isolation levels, weak consistency, reduction, refinement, formal verification.}

\maketitle

\section{Introduction}

\begin{figure}[t]
\centering
\begin{tikzpicture}[
  >={Stealth[length=1.6mm]},
  lv/.style={draw, rounded corners, font=\scriptsize, inner sep=2.5pt,
             minimum height=5mm, minimum width=8mm, align=center},
  red/.style={lv, fill=blue!12},
  nored/.style={lv, fill=violet!12}
]
\node[red] (RA) {Read\\ Atomicity\\ (RA)};
\node[nored, above right=2mm and -6mm of RA] (UA) {Update Atomicity\\ (UA)};
\node[red, below right=2mm and -6mm of RA] (CC) {Causal Consistency\\ (CC)};
\node[nored, right=4mm of RA] (PSI) {Parallel\\ Snapshot\\ Isolation (PSI)};
\node[nored, right=4mm of PSI] (SI) {Snapshot\\ Isolation (SI)};
\node[nored, right=4mm of SI] (SER) {Serializability\\ (SER)};
\node[nored, right=4mm of SER] (SSER) {Strict\\ SER\\ (SSER)};
\draw[->] (RA) -- (UA); \draw[->] (RA) -- (CC);
\draw[->] (UA) -- (PSI); \draw[->] (CC) -- (PSI);
\draw[->] (PSI) -- (SI); \draw[->] (SI) -- (SER); \draw[->] (SER) -- (SSER);
\end{tikzpicture}
\caption{Isolation levels, weaker ($\to$) stronger. Blue levels (IL $\leq$ CC) require a reduction step in addition to refinement. Violet levels (IL $\geq$ UA) directly refine the abstract model.}
\Description{Diagram showing isolation levels from weaker to stronger, with blue and violet regions indicating different model requirements.}
\label{fig:lattice}
\vspace{-3ex}
\end{figure}
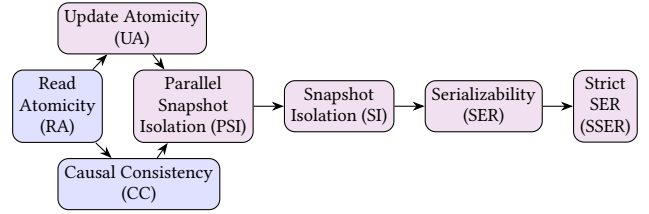

Modern databases trade strong isolation guarantees, such as serializability, for performance, offered by \emph{weak} isolation levels~\cite{Critique:SIGMOD1995,hat}, including read atomicity, causal consistency, snapshot isolation, and more (cf.~Figure~\ref{fig:lattice}). Weaker isolation levels permit more complex concurrent behavior, which makes the concurrency control protocols at those levels harder to get right. Indeed, weak-isolation bugs, stemming especially from design-level defects, have repeatedly surfaced in carefully designed and extensively tested production databases~\cite{Plume,jepsen-analyses}.

Over the past decades, sustained efforts have been made to develop reliable databases that support weakly isolated distributed transactions. Reasoning about isolation guarantees in these systems requires a formal semantics for isolation levels and a methodology for their formal verification.

%\nobi{I rewrote this paragraph:}
Although the semantics
have been extensively studied,
including
dependency graphs~\cite{bernstein1981concurrency}, 
abstract executions~\cite{DBLP:conf/concur/Cerone0G15}, 
and
operational semantics~\cite{DBLP:conf/ecoop/XiongCRG19},
tool support for verifying them remains rare.
%Prior work has proposed semantics, including dependency graphs~\cite{bernstein1981concurrency}, abstract executions~\cite{DBLP:conf/concur/Cerone0G15}, and
%operational semantics for isolation guarantees~\cite{DBLP:conf/ecoop/XiongCRG19}.
%
%\chsp{(Could possibly state here that verification tools for IL verification are rare.)}
%Although the semantics %are well-studied, 
%have been extensively studied for decades,
%tool support for verifying them remains rare.
%The recent VerIso~\cite{VerIso} framework mechanizes the operational semantics from~\cite{DBLP:conf/ecoop/XiongCRG19} in Isabelle/HOL~\cite{DBLP:books/sp/NipkowPW02}. 
Our recent VerIso~\cite{VerIso} framework
takes an important step toward closing this gap by mechanizing the operational semantics of~\cite{DBLP:conf/ecoop/XiongCRG19} in Isabelle/HOL~\cite{DBLP:books/sp/NipkowPW02}. 
%
%The framework is based on an abstract centralized model ($CIL$), which represents a multi-versioned key-value store and different client views.
Its core abstraction, 
namely the centralized isolation-level model ($CIL$), captures a 
multi-versioned key–value store and different client views. 
Parameterized by isolation levels, 
$CIL$ can be instantiated with a wide range of them
shown in Figure~\ref{fig:lattice}.
%It is parameterized by an isolation level (IL) and can be instantiated with any of the well-known 
%\nobi{``with a wide range of'' (not sure if everyone is well-known...) }
%levels shown in Figure~\ref{fig:lattice}. 
%VerIso establishes read atomicity (RA) as the model's baseline isolation guarantee by requiring view atomicity, i.e., a client either sees all versions written by a transaction or none. \chsp{possibly drop prev sentence from intro?}

In VerIso, one verifies isolation guarantees by showing that a concurrency control protocol \emph{refines} the abstract isolation-level model $CIL$ instantiated to the target level.
%VerIso uses \emph{refinement} as the standard technique for rigorously verifying isolation guarantees of concurrency control protocols. More precisely, a protocol is verified to satisfy an isolation guarantee by \emph{refining} the abstract isolation-level model instantiated to the target level. \chsp{could merge prev 2 sentences}
%VerIso has been used to verify that the two-phase commit protocol satisfies serializability~\cite{VerIso} and that the EigerPort+ protocol satisfies causal consistency~\cite{GhasemiradSprengerLiu+-TACAS25}.
%
%%%%\nobi{summarizing the challenges for making VerIso widely applicable/adopted. Then, subsequently, you could have paragraphs of your solutions that correspond to these challenges. In particular, it would also be better to start each solution paragraph with an "insight" sentence}
%
However, such proofs are labor-intensive 
%proofs (by Nobi)
and do not scale for two reasons: 
(1) there is a substantial abstraction gap between the abstract $CIL$ model and concrete protocol models, which requires rather involved refinement mappings linking the fine-grained protocol steps to the one-shot commit event in the $CIL$ model, and 
(2) refinement alone is sometimes insufficient (since there is no trace inclusion) and must be complemented by a Lipton-style \emph{reduction}~\cite{Lipton}. These reduction proofs are often done ad-hoc and on a per-protocol basis, which requires considerable additional effort.
%
%for some isolation levels refinement is insufficient (because there is no trace inclusion) and needs to be complemented by a Lipton-style reduction~\cite{Lipton}.
%
%For example, the causal consistency proof for EigerPort+~\cite{GhasemiradSprengerLiu+-TACAS25} requires a rather involved refinement mapping because of the much finer granularity of the protocol steps, compared to the one-shot commit event in the CIL model. Moreover, the reduction proof required in this case is done ad-hoc and requires a considerable additional effort. 
%\chsp{These problems are not limited to EP+.}

%However, for weaker isolation levels, refinement alone is often insufficient and a complementary \emph{reduction} technique~\cite{Lipton} is needed first to reduce concurrency while \emph{preserving the set of reachable states}. The reduction yields a more restricted model that can then be shown to refine an abstract isolation-level model. State-of-the-art deductive proofs, such as the recent verification of the causally consistent (CC) protocol Eiger-PORT+~\cite{EPplus}, implement this reduction step in an ad-hoc manner for each protocol, which is labor-intensive and does not scale in practice.

%\chsp{OUTLINE FOR REST: \\
%- We build a general reduction framework. \\
%- We analyze for which ILs reduction is required; identify two IL classes. \\
%- We build a hierarchy of models, with an abstract model for each class, once-and-for-all reduction, abstract protocol models.
%}

%\looseness=-1
In this work, we address these problems in three ways. %\nobi{four? DTXM counts, right? I tend to believe so because it stands as a different perspective (from system arch) to tackle the problem}
First, we introduce a general Lipton-style~\cite{Lipton} \emph{reduction} framework, whose application substantially reduces the effort needed for reduction proofs.
%for verifying weak isolation guarantees. \chsp{It is more general than for IL verification.} 
It is parameterized on an event system $ES$
(e.g., distributed transaction systems), 
an event dependency relation $D$
(e.g., write-read dependencies~\cite{Adya}), 
and 
a desired (good) event order $G$
(e.g., the order induced by transaction timestamps). 
%\nobi{these examples are concretizing event systems in the context of the domain we are working with }
Assuming that these relations satisfy three general conditions, including the linearity of $G$ and the commutativity of independent events, we prove that any execution of $ES$ is transformable into a $G$-ordered execution by commuting a finite number of independent event pairs (w.r.t.~$D$) while preserving the reachable states. To apply our result, one only needs to define the three parameters and discharge the three conditions. 
%We have fully mechanized this general result in Isabelle/HOL;
Notably, this general result is not limited to the database context.

%\nobi{have inline connections to databases: briefly explain what that means in the context of databases. defer this generality to the end of Intro to emphasize it, along with another point.}

%old: Second, to better understand the root cause for the need of reduction, we investigate isolation levels and related protocol correctness proofs requiring reduction (e.g.,~\cite{EPplus}).
Second, to gain deeper insight into why reduction is needed, we examine isolation levels and the associated protocol correctness proofs that require reduction (e.g.,~\cite{EPplus}).
%
%although the reduction framework radically shortens the reduction proof, proving a protocol correct still requires a per-protocol application of the framework. To address this, we investigate the lattice of seven isolation levels from Figure~\ref{fig:lattice} as well as protocol correctness proofs requiring reduction, to identify the root cause for the need of reduction and the levels where it is needed. %We then factor out the reduction for those levels and prove it once and for all.
%
We observe that write conflicts in protocols cause a mismatch between the logical order of events (e.g., according to timestamps) and their execution order, thereby necessitating a reduction proof to correct the mismatch. Since update atomicity (UA)~\cite{DBLP:conf/concur/Cerone0G15} is the weakest level preventing write conflicts, direct refinement alone suffices to prove correctness for this and all stronger levels. 
%Based on this observation, 
Accordingly, we classify the isolation levels in our lattice from Figure~\ref{fig:lattice} into two groups. The levels IL $\geq$ UA can be verified without reduction, whereas the verification of the two weaker levels IL $\leq$ CC, namely RA and CC, requires a combination of reduction and refinement. %Note that this result, about write conflicts, is not limited to these levels and is reflected in current prevalent isolation levels. \chsp{Not sure which (other?) current prevalent ILs you mean.}

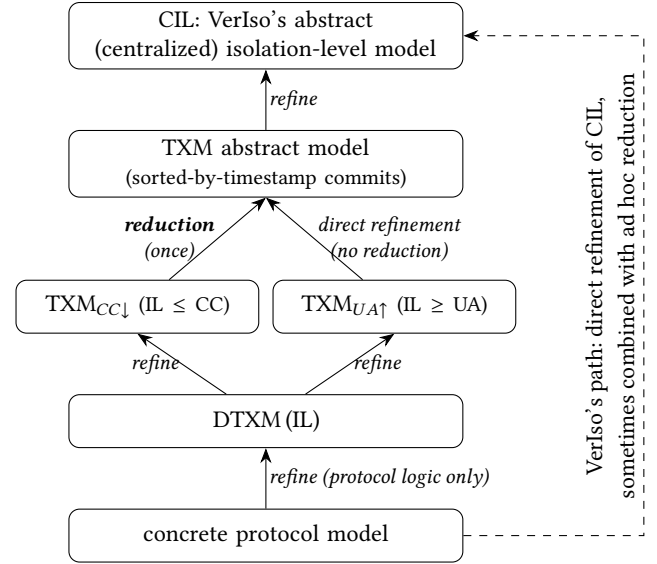
\begin{figure}[t]
\centering
\begin{tikzpicture}[
  >={Stealth[length=2mm]},
  box/.style={draw, rounded corners, align=center, font=\small,
              inner sep=3pt, minimum height=7mm, text width=50mm},
  lab/.style={font=\footnotesize\itshape, inner sep=1pt}
]
\node[box] (P) {concrete protocol model};
\node[box, above=8mm of P] (D) {DTXM\,(IL)};
\node[box, text width=30mm, above=8mm of D, xshift=-17mm] (Tcc) {TXM$_{CC\downarrow}$ \footnotesize{(IL $\leq$ CC)}};
\node[box, text width=30mm, above=8mm of D, xshift=17mm] (Tua) {TXM$_{UA\uparrow}$ \footnotesize{(IL $\geq$ UA)}};
\node[box, above=26mm of D] (R) {TXM abstract model\\\footnotesize{(sorted-by-timestamp commits)}};
\node[box, above=8mm of R] (V) {CIL: VerIso's abstract\\ (centralized) isolation-level model};
\draw[->] (P) -- node[lab,right]{refine (protocol logic only)} (D);
\draw[->] (D) -- node[lab,left,align=center]{refine} (Tcc.south);
\draw[->] (D) -- node[lab,right,align=center]{refine} (Tua.south);
\draw[->] (Tcc) -- node[lab,left,align=center]{\textbf{reduction}\\(once)} (R.south);
\draw[->] (Tua) -- node[lab,right,align=center]{direct refinement\\(no reduction)} (R.south);
\draw[->] (R) -- node[lab,right]{refine} (V);
\draw[dashed, ->] (P.east) -| (5,0) |- (V.east) node[pos=0.25, rotate=90, above, align=center]{\small{VerIso's path: direct refinement of CIL,}\\\small{sometimes combined with ad hoc reduction}};
\end{tikzpicture}
\caption{A hierarchy of models showing a transitive \emph{inclusion of reachable states} (via refinement or reduction) of a concrete protocol in the states permitted by its target IL. }
%\caption{The refinement stack. A concrete protocol refines DTXM, which refines TXM. The reduction is discharged once on the IL $\leq$ CC side; the IL $\geq$ UA side needs none. Above the TXM abstract model, the chain to VerIso's isolation-level model is proved once per level.}
\Description{Diagram showing the refinement stack: a concrete protocol refines DTXM, which refines TXM, with reductions discharged at specific points.}
\label{fig:stack}
\vspace{-1ex}
\end{figure}

Third, based on this insight, we construct a hierarchy of new abstract models with the aim of replacing per-protocol reduction proofs by a once-and-for-all reduction proof at a high abstraction level. Therefore, we construct a new abstract base model, TXM, based on transaction maps, which explicitly models timestamps and refines $CIL$. For this refinement to hold, we must restrict TXM such that the timestamp and execution order coincide.
From TXM, we then derive two more specialized abstract models, TXM$_{UA\uparrow}$ and TXM$_{CC\downarrow}$, one for each IL group (Figure~\ref{fig:stack}). 
We use our reduction framework to prove that TXM$_{CC\downarrow}$ reduces to TXM by reordering commits to respect the timestamp order. 
%, thus liberating protocol correctness proofs from bothering about the mismatch between timestamp and execution orders. 
We also show that TXM$_{UA\uparrow}$ directly refines TXM.
As a result, to prove a protocol's correctness, a refinement of the respective abstract model, instantiated to the desired IL, suffices.
%
%to obviate the need for per-protocol reduction proofs, we aim for a once-and-for-all reduction proof. To achieve this, we introduce, for each IL group, a new abstract model based on transaction maps, a structure that allows more concurrent behaviors than VerIso's $CIL$ model. These models are called TXM$_{UA\uparrow}$ and TXM$_{CC\downarrow}$ (Figure~\ref{fig:stack}). 
%We use our reduction framework to prove that TXM$_{CC\downarrow}$ reduces to a restricted transaction map model, TXM, where the timestamp and execution order coincide, and TXM, in turn, refines $CIL$. We also show that TXM$_{UA\uparrow}$ directly refines TXM. So, to prove a protocol's correctness, a refinement of the respective abstract model, instantiated to the desired IL, suffices.

%\looseness=-1
%\chsp{prev version:} For each group, we introduce a new intermediate abstract model based on transaction maps: TXM$_{UA\uparrow}$ and TXM$_{CC\downarrow}$. These are less restrictive than VerIso's $CIL$ model and allow more concurrent behaviors,\chsp{both of them?} thus serving as direct refinement targets for protocols. We prove a reduction, once and for all, from TXM$_{CC\downarrow}$ to a restricted model, TXM, that refines $CIL$. We also show that TXM$_{UA\uparrow}$ refines TXM and, by extension, $CIL$. As a result, subsequent protocol verifications inherit the reduction results, where applicable via TXM$_{CC\downarrow}$, and directly refine their corresponding abstract models.

Moreover, to factor out the verification effort of the shift from centralized to distributed settings, we introduce an abstract distributed protocol model (DTXM) parameterized over IL, which further refines the previous two models and captures the client/server two-phase commit (2PC) architecture shared by most modern distributed transaction systems. This concludes our hierarchy of intermediate models, which extends VerIso and reduces the verification of a new protocol to a single refinement of the distributed model 
%inheriting both the  2PC structure and the reduction for free 
(see 
Figure~\ref{fig:stack}).
%The models extend the VerIso framework, thereby reducing the abstraction gap between VerIso's abstract isolation-level model and concrete protocols, and also enable reasoning about different classes of isolation levels based on their need for reduction. 

\paragraph{Contributions} To summarize, this work makes the following contributions:
\begin{enumerate}
    \item A reduction framework, %mechanized in Isabelle/HOL, 
    parameterized on an event system, a dependency relation, and a desired event order, 
    %and mechanized in Isabelle/HOL, which can be applied to
    for verifying distributed transaction protocols with weak isolation guarantees (Section~\ref{sec:reduction}). %\chsp{tried to stress generality; however, unclear at this point why this is still needed for txn protocol verification after explanation of once-and-for-all reduction.}\shabnam{maybe we can skip talking about its generality here. We don't have enough case studies for it yet anyways. The one sentence in the intro body is enough I think.}
    \item A taxonomy of isolation levels ordered by whether their correctness proof follows from direct refinement of the abstract model (IL $\geq$ UA) or requires a \emph{reduction step} in addition (IL $\leq$ CC), which we factor out and prove using our reduction framework (Section~\ref{sec:taxonomy}).
    \item A hierarchy of abstract intermediate models that shrink the abstraction gap between VerIso's abstract model and concrete protocols (Sections~\ref{sec:txm} to \ref{sec:dtxm}).
\end{enumerate}

\noindent
The reduction framework, models' hierarchy, refinements, and the single reduction are all mechanized in Isabelle/HOL. %The DTXM refinement (Section~\ref{sec:dtxm}) is work in progress.
%toward the full version. %Section~\ref{sec:related} discusses related work.

% complements prior study on isolation levels
%\nobi{
This work also complements two lines of research.
First,  while prior work has primarily viewed
isolation levels through the lenses of system performance~\cite{hat,noc-noc,NOC:OSDI2020,9139787} and
black-box checking complexity~\cite{BiswasEnea19,10.1007/978-3-031-98685-7_15,10.1145/3742465}, 
we study them from the perspective of protocol verification and uncover a related dichotomy. % among them. \chsp{drop ``among them''?}
%}\shabnam{
Second, while prior reduction frameworks based on Lipton's movers target concurrent \emph{programs}, typically by inferring atomic blocks~\cite{CIVL15,Kragl18,Kragl20,ESOP26}, we provide a reduction framework for distributed systems that instead commutes protocol events
into a certain desired order.%}canonicalizes the order of events.
%Second, while prior work~\cite{}... we establish a more general reduction framework for reasoning about concurrent event systems.}

\section{A Reduction Framework in Isabelle/HOL}
\label{sec:reduction}

In this section, we introduce our general reduction framework, mechanized in Isabelle/HOL. We model database protocols and abstract transaction systems as labeled transition (event) systems. Our framework is based on the idea that two independent events in an execution \emph{commute}, i.e., swapping them disables neither event and yields the same final state.

Lipton in \cite{Lipton} uses this idea to group program statements into atomic blocks, to simplify reasoning about concurrent systems. We generalize this concept by commuting events into a given \emph{desired} order, while preserving reachable states. %This technique is quite general and not restricted to database protocols.
%\chsp{Not so clear whether we really generalize his approach: Can we always form ``atomic blocks'' of events using $D$ and $G$?}\shabnam{because it's "any order". As long as we can express the order that describes atomic blocks: like what you had done for the SER case}
 
%An event system has a set of states $S$ (with initial states $I \subseteq S$), and a set of events $E$, and a transition relation that defines the allowed transitions.
Our framework is parameterized on an event system $ES$, an event dependency relation $D$, and a partial function $G$ that specifies the desired (good) event order. $G$ assigns a linearly ordered label (e.g. a timestamp) to a subset of events. %whose order is of interest.

We identify a small set of assumptions on these parameters, which is required for the reduction to work:

\begin{enumerate}[leftmargin=15pt]
    \item $\forall\tau \in traces\;ES.\tau=\alpha \cdot e_1 \cdot e_2 \cdot \beta\;\land\;(e_1, e_2) \notin D \Rightarrow (e_2, e_1) \notin D$, i.e., if two events on the trace are independent, commuting them preserves their independence.
    \item $\forall\tau \in traces\;ES.\; \tau = \alpha \cdot e_2 \cdot \beta \cdot e_1 \cdot \gamma \land G(e_1) < G(e_2) \;\Rightarrow\;(e_2, e_1) \notin D$, i.e., "badly ordered" events are independent.
    \item $\forall\tau \in traces\;ES.\; \tau = \alpha \cdot e_1 \cdot \beta \cdot e_2 \cdot \gamma \land (e_1, e_2) \notin D \Rightarrow commute\;ES\;e_1\;e_2$, i.e., independent events commute.
\end{enumerate}

%To use the framework, one must instantiate it with an $ES$, a concrete dependency relation $D$ on its events that restricts commutes, and a desired order of events $G$, and prove these assumptions hold. The framework also introduces a well-founded measure that strictly decreases under commuting events towards a "sorted" execution, i.e., an execution where all events appear according to the desired order $G$. So the framework provides termination guarantees for the reduction process when all the above obligations are satisfied.

%\paragraph{Soundness.} The framework's reusable conclusion, proved once and generically, is:
\begin{theorem}[Reduction Soundness]
\label{thm:soundness}
Let $ES$ be an event system with an event dependency relation $D$ and a desired event order $G$, satisfying the framework's obligations. Then every reachable state of $ES$ is reachable by an execution where the event order is consistent with $G$ (a $G$-ordered execution).
\end{theorem}
\noindent The proof reorders an arbitrary execution into its sorted form by repeatedly swapping events. Then, a well-founded measure that strictly decreases under commuting events, marking progress towards a $G$-sorted execution, guarantees termination. Verifying an arbitrary-interleaving system thus reduces to verifying only its sorted executions.

\section{Abstract Base Model: Transaction Maps}
\label{sec:txm}

%\nobi{title is a bit odd} \shabnam{better? (old was: Transaction Maps Abstract Model)}

We introduce Transaction Maps (TXM), a new parameterized abstract model whose instantiations capture a range of well-known and prevalent isolation levels (Figure~\ref{fig:lattice}). The target isolation level $IL$ is specified by two model parameters: %, similar to VerIso's $CIL$ model:
\begin{itemize}[leftmargin=15pt]
\item $R_{IL}$, a relation on transaction IDs, and 
\item $\vShift_{IL}$, a predicate on allowed client view updates.
%about session guarantees (e.g. monotonic reads and read-your-writes). 
\end{itemize} We model TXM by defining its states and events as follows.

\paragraph{TXM states}
The TXM model's states $(M, U)$ consist of a \emph{transaction map} $M$ from transaction IDs to per-transaction states,
and a \emph{client views map} $U$ from client IDs to their views, i.e., the set of transactions they see. Unlike VerIso's $CIL$ model, which uses key-value version lists, this model stores information \emph{per transaction}, comprising (i) a read map, (ii) a write map, and (iii) a transaction timestamp.

\looseness=-1
This change reduces the abstraction gap to concrete protocols because one works directly with \emph{transaction IDs}, which serve both as the main way to access transaction states and as a replacement for list indices in client views. This approach can be seen as an operational version of abstract executions~\cite{DBLP:conf/concur/Cerone0G15}, as it defines a visibility relation between transactions via client views, and assigns timestamps to transactions, resembling the arbitration relation in abstract executions. TXM also offers view atomicity for free, i.e., clients see either all or none of other transactions' writes, and concrete protocols \emph{inherit} the atomic view by refining this model.

\paragraph{TXM events}
TXM has two events: a view extension event and an (atomic) commit event. The view extension event extends one client's view and preserves $M$. The commit event changes both $M$ and the committing client's view $u$ (in $U$), and requires that all of the following conditions hold for $M$, $u$, and their updates $M'$ and $u'$ resulting from the commit.
\begin{enumerate}
    \item $u$ is \emph{backward closed} under $R_{IL}$ (a.k.a. $\canCommit_{IL}$).
    \item The $\vShift_{IL}(M, u, M', u')$ predicate holds.
    \item The client reads the latest version in its view (\emph{last-write-wins} policy).
    \item The transaction ID and timestamp are fresh.
    \item For each involved write key, the timestamp is larger than all previous write transaction timestamps on that key (sorted-timestamps condition).
\end{enumerate}
The last condition restricts this model's executions to those in which the commits on each key are sorted by timestamp.

Finally, we show that this model refines %\footnote{Refinement is defined formally in the next section.}
VerIso's $CIL$ model. So, by transitivity, any model that refines TXM also refines $CIL$. (The reverse refinement, which would establish equivalence, is work in progress.)

\section{Isolation Levels Taxonomy, Refinement, and Once-and-for-All Reduction}
\label{sec:taxonomy}

%\david{This starts out technical. For each section I would bring the reader back into a higher-level story line so s/he knows why it is worth reading the section.}

In this section, we first discuss general proof techniques for transaction isolation verification. Then, we propose a new characterization of isolation levels and the associated proof techniques. 
In order to capture recurring proof patterns, we introduce two new models in our model hierarchy, namely TXM$_{CC\downarrow}$ and TXM$_{UA\uparrow}$ (see also Figure~\ref{fig:stack}) and generalize those patterns as a once-and-for-all effort.

%In this section, we develop a verification-oriented view of isolation levels based on inherent characteristics of these levels. We first review the general proof techniques available for transaction-isolation verification, then use them to characterize isolation levels by the technique each requires. To capture the recurring proof patterns this reveals, we introduce two new models into our hierarchy (Figure~\ref{fig:stack}) and discharge those patterns once and for all.

%r preserves initial states, and (r, \pi) maps ES1 transitions to ES2 transitions.

Refinement is a standard, commonly used technique to show that concurrency control protocols satisfy their isolation guarantees. A refinement is defined between two event systems at different abstraction levels, using a refinement mapping pair $(r, \pi)$ that maps the states and events of one system to those of the other, respectively. We say an event system $ES_1$ refines event system $ES_2$, if (i) $r$ preserves initial states, and (ii) $(r, \pi)$ maps $ES_1$ transitions to $ES_2$ transitions, i.e., if $s\myrightarrow{e}_1 s'$, then $ r(s)\myrightarrow{\pi(e)}_2 r(s')$.

Refinement guarantees inclusion of reachable states modulo~$r$, i.e., $r(reach(ES_1)) \subseteq reach(ES_2)$, with $r$ applied pointwise. For database protocols, showing that the concrete concurrency control protocol refines the desired isolation level's event system proves that the protocol's reachable states are included in those of the isolation level, thereby establishing satisfaction of that level's guarantees.

Note, however, that for such correctness proofs, a \emph{direct} refinement of the abstract isolation-level model (whether of $CIL$ or base TXM model) does not always work. We can observe this, for example, in the Eiger-PORT+\cite{EPplus} protocol. In this protocol, a read transaction always reads the version with the \emph{highest timestamp} below a certain global safe time (which determines a subset of already committed transactions). %(an underapproximation of the set of committed transactions).
In contrast, both abstract models $CIL$ and TXM require that a client always reads the \emph{latest} committed version in its view, which is determined by the \emph{execution order} of commit events. However, the execution order of commits and the \emph{order of the associated commit timestamps} may not coincide. We call such commits \emph{inverted commits}.

Inverted commits in an execution cannot simulate the behavior of the abstract isolation-level models $CIL$ or TXM. In $CIL$, simulating such commits requires \emph{inserting} a key’s new version to its version list rather than \emph{appending} it, which is not allowed. Similarly, in TXM, which refines $CIL$, such commits are excluded by the sorted-timestamps condition. We solve this problem by combining reduction and refinement. Reduction orders the commit events by their timestamps, resulting in the inversion-free (a.k.a. restricted/sorted) model TXM, and TXM subsequently refines the $CIL$ abstract model.

To pinpoint the root cause of this problem, we examined a range of isolation levels and observed that if a level disallows \emph{write conflicts}, then this kind of timestamp inversion is avoided. Based on this observation, we classify isolation levels into two groups, as shown in Figure~\ref{fig:lattice}.

\paragraph{$\mathbf{TXM_{UA\uparrow}}$ and direct refinement}
For levels stronger than update atomicity, \textbf{IL $\geq$ UA} (UA, PSI, SI, SER, SSER), we introduce the TXM$_{UA\uparrow}$ intermediate model, which differs from the TXM abstract model in two ways.

First, IL $\geq$ UA entails that IL's commit condition implies the UA commit condition ($\canCommit_{UA}$). This condition requires the client view to be backward closed under the relation $R_{UA}=\bigcup_{(k, v) \in write\_map} WW^{-1}_M(k)$, meaning it should include \emph{all} versions of the keys in the transaction's write map (because of closure on write-write dependency), essentially avoiding write conflicts.

Second, the model replaces the restrictive \emph{sorted timestamps} condition of TXM with a much weaker timestamp condition, namely requiring a client to choose a timestamp greater than the transactions' timestamps \emph{in its view}. This condition is respected by timestamped protocols, since the timestamp always increases locally for a client, and its choice depends only on the client's view, rather than on \emph{all} committed transactions of the involved write keys. 

Combining these two conditions, we prove that for all keys a transaction commits to, the timestamp will be greater than those of the existing transactions in the client's view, which must, in turn, include \emph{all} versions of those keys. Therefore, this model directly refines the TXM abstract model, and \emph{no} reduction is needed to put commits in the correct order.

\paragraph{$\mathbf{TXM_{CC\downarrow}}$ and the once-and-for-all reduction}
For levels \textbf{IL $\leq$ CC} (RA and CC), we introduce the TXM$_{CC\downarrow}$ intermediate model, which removes the \emph{sorted timestamps} condition of TXM in exchange for a few weaker timestamp conditions similar to TXM$_{UA\uparrow}$. Here, however, the instantiated levels are weaker and therefore permit more complex concurrent behavior. This breaks trace inclusion between this model and TXM, making a direct refinement of TXM infeasible.
%However, since the instantiated levels here are weaker, more complex concurrent behavior is allowed which means there is no trace inclusion between this model and $TXM$, and thus a direct refinement of TXM becomes infeasible. \chsp{why? no trace inclusion (if space permits)}

Therefore, to prove the inclusion of the set of reachable states for the IL $\leq$ CC region, represented by the TXM$_{CC\downarrow}$ model, we instantiate the framework from Section~\ref{sec:reduction} a \emph{single}
time, with an event dependency relation $D_{CC}$ and the commit timestamp order $G_{CC}$, as the desired good order.

$D_{CC}$ defines two events $e_1$ and $e_2$ as dependent in two cases: (1) if both events are from the same client, or (2) if $e_1$ is a commit, and its transaction ID is in $e_2$'s client view (capturing write-read dependency). 
%of the same client as dependent. Additionally, if a commit's transaction ID appears in the view of a later commit or view extension event, then the latter is dependent on the first commit, capturing write-read dependency.

\looseness=-1
$G_{CC}$ maps commit events to their timestamps, thereby indicating the timestamp order as the desired order of commits.

The instantiation yields three proof obligations (see Section~\ref{sec:reduction}), which we prove hold for TXM$_{CC\downarrow}$, using $D_{CC}$ and $G_{CC}$. As a result, TXM$_{CC\downarrow}$ is proven reducible to the TXM model,
in which the commits are ordered by their timestamps.
Thus, every concrete protocol that refines TXM$_{CC\downarrow}$ automatically satisfies the reduced semantics and
the level guarantee, with no further reduction reasoning. The reduction argument, the hardest part of the proof, is paid for exactly once. 

%Figure~\ref{fig:stack} shows the architecture of our models and composed refinements and reduction. %\chsp{I would place reference to Fig 2 much earlier, as an overview/map to follow.}

\section{Distributed Transaction Maps}
\label{sec:dtxm}

Distribution in most concurrency control protocols is structured as clients and servers running a two-phase commit protocol. We capture
this in the Distributed Transaction Maps (DTXM) model, %(\code{Txm\_CCd\_Dist.thy})
which bridges the centralized TXM models to distributed protocol models.

DTXM's global state includes client and server configurations, and a map of transactions to timestamps. A client configuration stores the client's view, its current transaction's ID, timestamp, and state machine $\mathit{init}$$\to$$\mathit{ops}$$\to$$\mathit{prepared}$$\to$$\mathit{committed}/\mathit{aborted}$. A server configuration stores, for each ($t$, $k$) transaction ID and key pair, 
%(considering key-to-server assignments), 
a state machine $\mathit{idle}$$\to$$\mathit{read}/\mathit{prepared}$$\to$$\mathit{committed}/\mathit{aborted}$. DTXM events form a full 2PC protocol: read request and response, prepare, commit, and abort.

We show that DTXM refines previous TXM models by proving that its client commit event refines the \emph{one-shot atomic commit} of TXM models. %(\code{txm\_cc\_dist\_refines\_txm\_cc})
%is currently being mechanized. The intended consequence is that
Therefore, a future protocol correctness proof requires \emph{only a refinement} of DTXM.

\section{Conclusions and Future Work}
We have introduced a new approach to transaction isolation verification and factored out its labor-intensive proof steps by developing a generic reduction framework and a hierarchy of abstract models. We have also shed new light on isolation levels from a protocol verification perspective by distinguishing those that can be verified by refinement alone from those that additionally require reduction.

In order to validate our approach at both ends of the isolation-level lattice, we are currently verifying two concurrency control protocols: for the IL $\leq$ CC class, we reverify Eiger-PORT+\cite{EPplus} as causally consistent, and for the IL $\geq$ UA class, we verify that ROLA~\cite{rola} satisfies UA. The reverification highlights the reduced proof effort as a result of using our reduce-once model hierarchy, while the second case study witnesses the generality of our approach.

\clearpage
\bibliographystyle{ACM-Reference-Format}
\bibliography{SCCP}

@article{VerIso,
  author    = {Shabnam Ghasemirad and Si Liu and Christoph Sprenger and Luca Multazzu and David Basin},
  title     = {{VerIso}: Verifiable Isolation Guarantees for Database Transactions},
  journal   = {Proc. {VLDB} Endow.},
  volume    = {18},
  number    = {5},
  pages     = {1362--1375},
  year      = {2025},
  doi       = {10.14778/3718057.3718065}
}

@inproceedings{EPplus,
  author    = {Shabnam Ghasemirad and Christoph Sprenger and Si Liu and Luca Multazzu and David Basin},
  title     = {Pushing the Limit: Verified Performance-Optimal Causally-Consistent Database Transactions},
  booktitle = {Tools and Algorithms for the Construction and Analysis of Systems (TACAS), ETAPS},
  series    = {LNCS},
  volume    = {15698},
  pages     = {43--62},
  year      = {2025},
  doi       = {10.1007/978-3-031-90660-2\_3},
}

@article{Lipton,
	author = {Richard J. Lipton},
	title = {Reduction: {A} Method of Proving Properties of Parallel Programs},
	journal = {Commun. {ACM}},
	volume = {18},
	number = {12},
	year = {1975},
	pages = {717--721},
    url          = {https://doi.org/10.1145/361227.361234},
    doi          = {10.1145/361227.361234}
}

@inproceedings{CIVL15,
  author    = {Chris Hawblitzel and Erez Petrank and Shaz Qadeer and Serdar Tasiran},
  title     = {Automated and Modular Refinement Reasoning for Concurrent Programs},
  booktitle = {Computer Aided Verification},
  series    = {LNCS},
  volume    = {9207},
  pages     = {449--465},
  year      = {2015},
  doi       = {10.1007/978-3-319-21668-3\_26},
}

@inproceedings{Kragl18,
  author    = {Bernhard Kragl and Shaz Qadeer},
  title     = {Layered Concurrent Programs},
  booktitle = {Computer Aided Verification},
  series    = {LNCS},
  volume    = {10981},
  pages     = {79--102},
  year      = {2018},
  doi       = {10.1007/978-3-319-96145-3\_5},
}

@inproceedings{Kragl20,
  author    = {Bernhard Kragl and Shaz Qadeer and Thomas A. Henzinger},
  title     = {Refinement for Structured Concurrent Programs},
  booktitle = {Computer Aided Verification},
  series    = {LNCS},
  volume    = {12224},
  pages     = {275--298},
  year      = {2020},
  doi       = {10.1007/978-3-030-53288-8\_14},
}

@inproceedings{ESOP26,
author = {Gangamreddypalli, Namratha and Enea, Constantin and Qadeer, Shaz},
title = {Reduction for Structured Concurrent Programs},
year = {2026},
isbn = {978-3-032-22719-5},
publisher = {Springer-Verlag},
address = {Berlin, Heidelberg},
doi = {10.1007/978-3-032-22720-1_10},
booktitle = {Programming Languages and Systems: 35th European Symposium on Programming, ESOP 2026, Held as Part of the International Joint Conferences on Theory and Practice of Software, ETAPS 2026, Turin, Italy, April 11–16, 2026, Proceedings, Part I},
pages = {252–282},
numpages = {31},
location = {Turin, Italy}
}

@article{Plume,
  author       = {Si Liu and
                  Long Gu and
                  Hengfeng Wei and
                  David Basin},
  title        = {Plume: Efficient and Complete Black-Box Checking of Weak Isolation
                  Levels},
  journal      = {Proc. {ACM} Program. Lang.},
  volume       = {8},
  number       = {{OOPSLA}},
  pages        = {876--904},
  year         = {2024},
  url          = {https://doi.org/10.1145/3689742},
  doi          = {10.1145/3689742}
}

@inproceedings{BiswasEnea19,
  author    = {Ranadeep Biswas and Constantin Enea},
  title     = {On the Complexity of Checking Transactional Consistency},
  booktitle = {Proc. {ACM} Program. Lang. (OOPSLA)},
  volume    = {3},
  pages     = {165:1--165:28},
  year      = {2019},
  doi       = {10.1145/3360591},
}

@article{noc-noc,
author = {Liu, Si and Multazzu, Luca and Wei, Hengfeng and Basin, David A.},
title = {NOC-NOC: Towards Performance-optimal Distributed Transactions},
year = {2024},
issue_date = {February 2024},
publisher = {ACM},
volume = {2},
number = {1},
url = {https://doi.org/10.1145/3639264},
doi = {10.1145/3639264},
journal = {Proc. ACM Manag. Data},
month = mar,
articleno = {9},
numpages = {25}
}

@article{hat,
  author       = {Peter Bailis and
                  Aaron Davidson and
                  Alan D. Fekete and
                  Ali Ghodsi and
                  Joseph M. Hellerstein and
                  Ion Stoica},
  title        = {Highly Available Transactions: Virtues and Limitations},
  journal      = {Proc. {VLDB} Endow.},
  volume       = {7},
  number       = {3},
  pages        = {181--192},
  year         = {2013},
  url          = {http://www.vldb.org/pvldb/vol7/p181-bailis.pdf},
  doi          = {10.14778/2732232.2732237},
  bibsource    = {dblp computer science bibliography, https://dblp.org}
}

@article{Critique:SIGMOD1995,
	author     = {Berenson, Hal and Bernstein, Phil and Gray, Jim and Melton, Jim and O'Neil, Elizabeth and O'Neil, Patrick},
	doi        = {10.1145/568271.223785},
	issue_date = {May 1995},
	journal    = {SIGMOD Rec.},
	month      = may,
	number     = {2},
	numpages   = {10},
	pages      = {1–10},
	title      = {A critique of ANSI SQL isolation levels},
	volume     = {24},
	year       = {1995}
}

@inproceedings{DBLP:conf/ecoop/XiongCRG19,
  author       = {Shale Xiong and
                  Andrea Cerone and
                  Azalea Raad and
                  Philippa Gardner},
  title        = {Data Consistency in Transactional Storage Systems: {A} Centralised
                  Semantics},
  booktitle    = {34th European Conference on Object-Oriented Programming, {ECOOP} 2020},
  series       = {LIPIcs},
  volume       = {166},
  pages        = {21:1--21:31},
  publisher    = {Schloss Dagstuhl - Leibniz-Zentrum f{\"{u}}r Informatik},
  year         = {2020},
  url          = {https://doi.org/10.4230/LIPIcs.ECOOP.2020.21},
  doi          = {10.4230/LIPICS.ECOOP.2020.21}
}

@phdthesis{Adya,
  title={Weak consistency: a generalized theory and optimistic implementations for distributed transactions},
  author={Adya, Atul},
  year={1999},
  school={Massachusetts Institute of Technology, Department of Electrical Engineering and Computer Science}
}

@inproceedings{psi,
  author       = {Yair Sovran and
                  Russell Power and
                  Marcos K. Aguilera and
                  Jinyang Li},
  editor       = {Ted Wobber and
                  Peter Druschel},
  title        = {Transactional storage for geo-replicated systems},
  booktitle    = {Proceedings of the 23rd {ACM} Symposium on Operating Systems Principles
                  2011, {SOSP} 2011},
  pages        = {385--400},
  publisher    = {{ACM}},
  year         = {2011},
  url          = {https://doi.org/10.1145/2043556.2043592},
  doi          = {10.1145/2043556.2043592}
}

@inproceedings{si,
  author       = {Hal Berenson and
                  Philip A. Bernstein and
                  Jim Gray and
                  Jim Melton and
                  Elizabeth J. O'Neil and
                  Patrick E. O'Neil},
  editor       = {Michael J. Carey and
                  Donovan A. Schneider},
  title        = {A Critique of {ANSI} {SQL} Isolation Levels},
  booktitle    = {Proceedings of the 1995 {ACM} {SIGMOD} International Conference on Management of Data},
  pages        = {1--10},
  publisher    = {{ACM} Press},
  year         = {1995},
  url          = {https://doi.org/10.1145/223784.223785},
  doi          = {10.1145/223784.223785}
}

@book{DBLP:books/sp/NipkowPW02,
	author = {Tobias Nipkow and Lawrence C. Paulson and Markus Wenzel},
	bibsource = {dblp computer science bibliography, https://dblp.org},
	doi = {10.1007/3-540-45949-9},
	isbn = {3-540-43376-7},
	publisher = {Springer},
	series = {Lecture Notes in Computer Science},
	title = {Isabelle/HOL - {A} Proof Assistant for Higher-Order Logic},
	url = {https://doi.org/10.1007/3-540-45949-9},
	volume = {2283},
	year = {2002}}

@inproceedings{DBLP:conf/concur/Cerone0G15,
  author       = {Andrea Cerone and
                  Giovanni Bernardi and
                  Alexey Gotsman},
  editor       = {Luca Aceto and
                  David de Frutos{-}Escrig},
  title        = {A Framework for Transactional Consistency Models with Atomic Visibility},
  booktitle    = {26th International Conference on Concurrency Theory, {CONCUR} 2015},
  series       = {LIPIcs},
  volume       = {42},
  pages        = {58--71},
  publisher    = {Schloss Dagstuhl - Leibniz-Zentrum f{\"{u}}r Informatik},
  year         = {2015},
  url          = {https://doi.org/10.4230/LIPIcs.CONCUR.2015.58},
  doi          = {10.4230/LIPICS.CONCUR.2015.58}
}

@article{bernstein1981concurrency,
  title={Concurrency control in distributed database systems},
  author={Bernstein, Philip A and Goodman, Nathan},
  journal={ACM Computing Surveys (CSUR)},
  volume={13},
  number={2},
  pages={185--221},
  year={1981},
  publisher={ACM New York, NY, USA},
  url          = {https://doi.org/10.1145/356842.356846},
  doi          = {10.1145/356842.356846}
}

@inproceedings{NOC:OSDI2020,
    author = {Haonan Lu and Siddhartha Sen and Wyatt Lloyd},
    title = {{Performance-Optimal} {Read-Only} Transactions},
    booktitle = {14th USENIX Symposium on Operating Systems Design and Implementation (OSDI 20)},
    year = {2020},
    isbn = {978-1-939133-19-9},
    pages = {333--349},
    url = {https://www.usenix.org/conference/osdi20/presentation/lu},
    publisher = {USENIX Association},
    month = nov
}

@article{ua,
  author       = {Si Liu and
                  Peter Csaba {\"{O}}lveczky and
                  Qi Wang and
                  Indranil Gupta and
                  Jos{\'{e}} Meseguer},
  title        = {Read atomic transactions with prevention of lost updates: {ROLA} and
                  its formal analysis},
  journal      = {Formal Aspects Comput.},
  volume       = {31},
  number       = {5},
  pages        = {503--540},
  year         = {2019},
  url          = {https://doi.org/10.1007/s00165-019-00489-w},
  doi          = {10.1007/S00165-019-00489-W}
}

@misc{jepsen-analyses,
	author = {Jepsen},
	title = {{Jepsen Analyses}},
    year = {2026},
	url = {https://jepsen.io/analyses},
    lastaccessed = {June~1, 2026}
}

@article{rola,
  author       = {Si Liu and
                  Peter Csaba {\"{O}}lveczky and
                  Qi Wang and
                  Indranil Gupta and
                  Jos{\'{e}} Meseguer},
  title        = {Read atomic transactions with prevention of lost updates: {ROLA} and
                  its formal analysis},
  journal      = {Formal Aspects Comput.},
  volume       = {31},
  number       = {5},
  pages        = {503--540},
  year         = {2019},
  url          = {https://doi.org/10.1007/s00165-019-00489-w},
  doi          = {10.1007/S00165-019-00489-W}
}

@inproceedings{9139787,
  author={Antoniadis, Karolos and Didona, Diego and Guerraoui, Rachid and Zwaenepoel, Willy},
  booktitle={2020 IEEE International Parallel and Distributed Processing Symposium (IPDPS)}, 
  title={The Impossibility of Fast Transactions}, 
  year={2020},
  volume={},
  number={},
  pages={1143-1154},
  doi={10.1109/IPDPS47924.2020.00120}
}

@inproceedings{10.1007/978-3-031-98685-7_15,
    author = {Bouajjani, Ahmed and Enea, Constantin and Rom\'{a}n-Calvo, Enrique},
    title = {On the Complexity of Checking Mixed Isolation Levels for SQL Transactions},
    year = {2025},
    isbn = {978-3-031-98684-0},
    publisher = {Springer-Verlag},
    url = {https://doi.org/10.1007/978-3-031-98685-7_15},
    doi = {10.1007/978-3-031-98685-7_15},
    booktitle = {Computer Aided Verification: 37th International Conference, CAV 2025, Zagreb, Croatia, July 23-25, 2025, Proceedings, Part IV},
    pages = {315–337},
    numpages = {23}
}

@article{10.1145/3742465,
    author = {M\o{}ldrup, Lasse and Pavlogiannis, Andreas},
    title = {AWDIT: An Optimal Weak Database Isolation Tester},
    year = {2025},
    issue_date = {June 2025},
    publisher = {ACM},
    volume = {9},
    number = {PLDI},
    url = {https://doi.org/10.1145/3742465},
    doi = {10.1145/3742465},
    journal = {Proc. ACM Program. Lang.},
    month = jun,
    articleno = {209},
    numpages = {25}
}

\appendix 
% appendix is allowed 
\end{document}